\documentstyle[12pt,epsfig,wrapfig]{article}
\newcommand{\bm}[1]{\mbox{\boldmath $#1$}}
\begin{document}
\begin{flushright}
NIKHEF 96-024\\
hep-ph/9610288
\end{flushright}
\begin{center}
{\large \bf
Quark transverse momentum in hard scattering processes
\footnote{Contributed paper at the 12$^{th}$ International Symposium on
High Energy Spin Physics, Amsterdam, Sept. 10-14, 1996}
}\\
\vspace{5mm}
R. Jakob$^1$, A. Kotzinian$^{2,3,4}$, \underline{P.J. Mulders}$^{1,5}$ and
J. Rodrigues$^{1,6}$
\\
\vspace{5mm}
{\small\it
(1) NIKHEF, P.O.Box 41882, 1009 DB Amsterdam, The Netherlands\\
(2) Yerevan Physics Institute, AM-375036 Yerevan, Armenia\\
(3) Universit\"at Mainz, D-55099 Mainz, Germany\\
(4) JINR, RU-141980 Dubna, Russia\\
(5) Free University, De Boelelaan 1081, 1081 HV Amsterdam, The Netherlands\\
(6) Instituto Superior T\'ecnico, Departamento de Fisica, 1096 Lisboa, Portugal
\\ }
\end{center}

\begin{center}
%
\begin{minipage}{130 mm}
\small
The role of transverse momentum of quarks in semi-inclusive
leptoproduction will be discussed.
It involves generalized distribution and fragmentation functions which
depend on both the longitudinal lightcone momentum fraction of the quarks
and on the transverse momentum. Constraints on these functions and relations 
between them arise as a consequence of the QCD equation of motion for the 
quark fields, Lorentz invariance and of C, P, and T invariance of the 
strong interactions. Experimentally one has access to the functions for 
instance in measurements of azimuthal asymmetries in hard processes.
\end{minipage}
\end{center}

Quark distribution functions and quark fragmentation functions appear in
the field-theoretical description of hard processes as the parts that
connect the quark and gluon lines to hadrons in the initial or final state.
These parts, containing the {\em soft} physics can be considered as 
(connected) matrix elements of nonlocal
operators built from quark and gluon fields [1]. The matrix elements
appearing in the leading order calculation for 1-particle inclusive
leptoproduction are
\begin{equation}
\Phi_{ij}(p,P,S) = \frac{1}{(2\pi)^4}\int d^4\xi\ e^{i\,p\cdot \xi}
\langle P,S \vert \overline \psi_j(0) \psi_i(\xi) \vert P,S \rangle,
\end{equation}
for the distribution part of quarks with momentum $p$ in the target hadron 
with momentum $P$ and spin $S$ and
\begin{equation}
\Delta_{ij}(k,P_h,S_h)  =  \sum_X \frac{1}{(2\pi)^4}\int d^4\xi
\ e^{ik\cdot \xi}\, \langle 0 \vert \psi_i(\xi) \vert P_h, X \rangle
\langle P_h,X \vert \overline \psi_j(0) \vert 0 \rangle, 
\end{equation}
describing the decay of a quark with momentum $k$ into a final state 
containing one specific hadron $h$ with momentum $P_h$.

In inclusive lepton-hadron scattering one finds in leading order of an
expansion in $1/Q$, where $-q^2 = Q^2$ is the spacelike momentum 
transfer squared, that the relevant matrix element at high energy is 
$\Phi(x)$ = $\int dp^-d^{\,2}p_T\,\Phi$,
in which case the nonlocality becomes a lightlike separation in $\xi^-$.
The lightlike $+$-direction is determined by the (up to mass effects) 
lightlike vector $P$.

The important extension when transverse momenta play a role, e.g. in processes
in which there are at least two hadrons in addition to the virtual photon
(or W/Z) such as 1-particle inclusive lepton-hadron scattering
($\ell H \rightarrow \ell^\prime h X$) or Drell-Yan ($A B \rightarrow
\mu^+ \mu^- X$), is the presence of transverse separations in the nonlocal
matrix elements.
Using constraints coming from hermiticity, parity and time reversal invariance
the most general parametrization of the Dirac projections,
\begin{equation}
\Phi^{[\Gamma]}(x,\bm p_T) = 
\frac{1}{2}\int dp^-\ Tr(\Phi\,\Gamma) =
\left. \int \frac{d\xi^-d^2\xi_T}{2\,(2\pi)^3} \
e^{ik\cdot \xi}
\,\langle P,S \vert \overline \psi (0) \Gamma 
\psi(\xi) \vert P,S \rangle \right|_{\xi^+ = 0} ,
\end{equation}
depending on $x = p^+/P^+$ and $\bm p_T$,
yields in leading order for a polarized spin 1/2 hadron [2] 
\begin{eqnarray}
& & \Phi^{[\gamma^+]} =
f_1(x ,\bm p_T) ,
\\ & & \Phi^{[\gamma^+ \gamma_5]} =
\lambda\,g_{1L}(x ,\bm p_T)
+ g_{1T}(x ,\bm p_T)\,\frac{(\bm p_T\cdot\bm S_T)}{M} ,
\\ & & \Phi^{[ i \sigma^{i+} \gamma_5]} =
S_T^i\,h_{1}(x ,\bm p_T)
+ \frac{\lambda p_T^i}{M}\,h_{1L}^\perp(x ,\bm p_T)
- \frac{\left(p_T^i p_T^j + \frac{1}{2}\bm p_T^2g_T^{ij}\right) S_{Tj}}{M^2}
\,h_{1T}^\perp(x ,\bm p_T),
\end{eqnarray}
where $\lambda$ and $S_T^\alpha$ are the (lightcone) helicity and the
transverse spin.
The $\bm p_T$-integrated results, relevant in inclusive scattering, are
the lightcone momentum distributions for unpolarized quarks, 
$\Phi^{[\gamma^+]}$ = $f_1(x)$, 
the quark helicity distribution 
$\Phi^{[\gamma^+\gamma_5]}$ = $\lambda\,g_1(x)$, and
the quark transverse spin distribution
$\Phi^{[i\sigma^{\alpha+}\gamma_5]}$ = $S_T^\alpha\,h_1(x)$. 

In measurements of azimuthal asymmetries one becomes sensitive to
$\bm p_T$-weighted functions $\Phi_{\partial^\alpha}^{[\Gamma]}(x)$ = 
$(1/2)\int dp^-d^{\,2}p_T\ p_T^\alpha\,Tr(\Phi\,\Gamma)$, e.g.
\begin{equation}
\frac{1}{M}\Phi_{\partial^\alpha}^{[\gamma^+\gamma_5]}(x)
= S_T^\alpha\,\int d^{\,2}p_T\,\frac{\bm p_T^2}{2M^2}\,g_{1T}(x,\bm p_T)
\equiv S_T^\alpha\,g_{1T}^{(1)}(x),
\end{equation}
first introduced (albeit with another name) in ref. [3].
These functions can be related to $\bm p_T$-averaged higher twist correlation
functions such as
\begin{equation}
\Phi^{[\gamma^\alpha\gamma_5]}(x) = \frac{M}{P^+}\,S_T^\alpha\,g_T(x).
\end{equation}
One has the relation $g_2(x)$ = $(g_T-g_1)(x)$ = $d\,g_{1T}^{(1)}/dx$.
The factor $M/P^+$ required by Lorentz invariance,
leads to a factor $M/Q$ in the cross sections. 

In order to obtain the cross section for a 1-particle inclusive process 
the distribution part must be combined with a fragmentation part. At
leading order for the case of summing over polarizations of the final state
hadron one encounters e.g.
\begin{equation}
\Delta^{[\gamma^-]}(z) = \frac{1}{4z}\,\int dk^+d^{\,2}\bm k_T\,Tr(\Delta
\gamma^-) = D_1(z),
\end{equation}
where $z$ = $P_h^-/k^-$.
An example of a nonvanishing azimuthal asymmetry in scattering polarized
leptons from transversely polarized nucleons is
\begin{eqnarray}
&&\int d^2\bm P_{h\perp}\,\frac{\vert \bm P_{h\perp}\vert}
{M\,z_h} \,cos(\phi_h-\phi_S)
\,\frac{d\sigma_{LT}}{dx_B\,dy\,dz_h\,d^2\bm P_{h\perp}}
\nonumber \\ &&\qquad\qquad \qquad\qquad
= \frac{2\pi \alpha^2\,s}{Q^4}\,\lambda_e\,\vert \bm S_T \vert
\,y(2-y)\sum_{a,\bar a} e_a^2
\,x_B\,g_{1T}^{(1)a}(x_B) D^a_1(z_h),
\end{eqnarray} 
where the azimuthal angles are defined with respect to the lepton scattering
plane.
In this expression we have used the usual scaling variables, $x_B$ =
$Q^2/2P\cdot q$, $y$ = $P\cdot q/P\cdot k$ and $z_h$ = $P\cdot P_h/P\cdot q$
and we have included the summation over quark flavors and the weighting
with the quark charges squared.

\begin{wrapfigure}{r}{6 cm}
\epsfig{figure=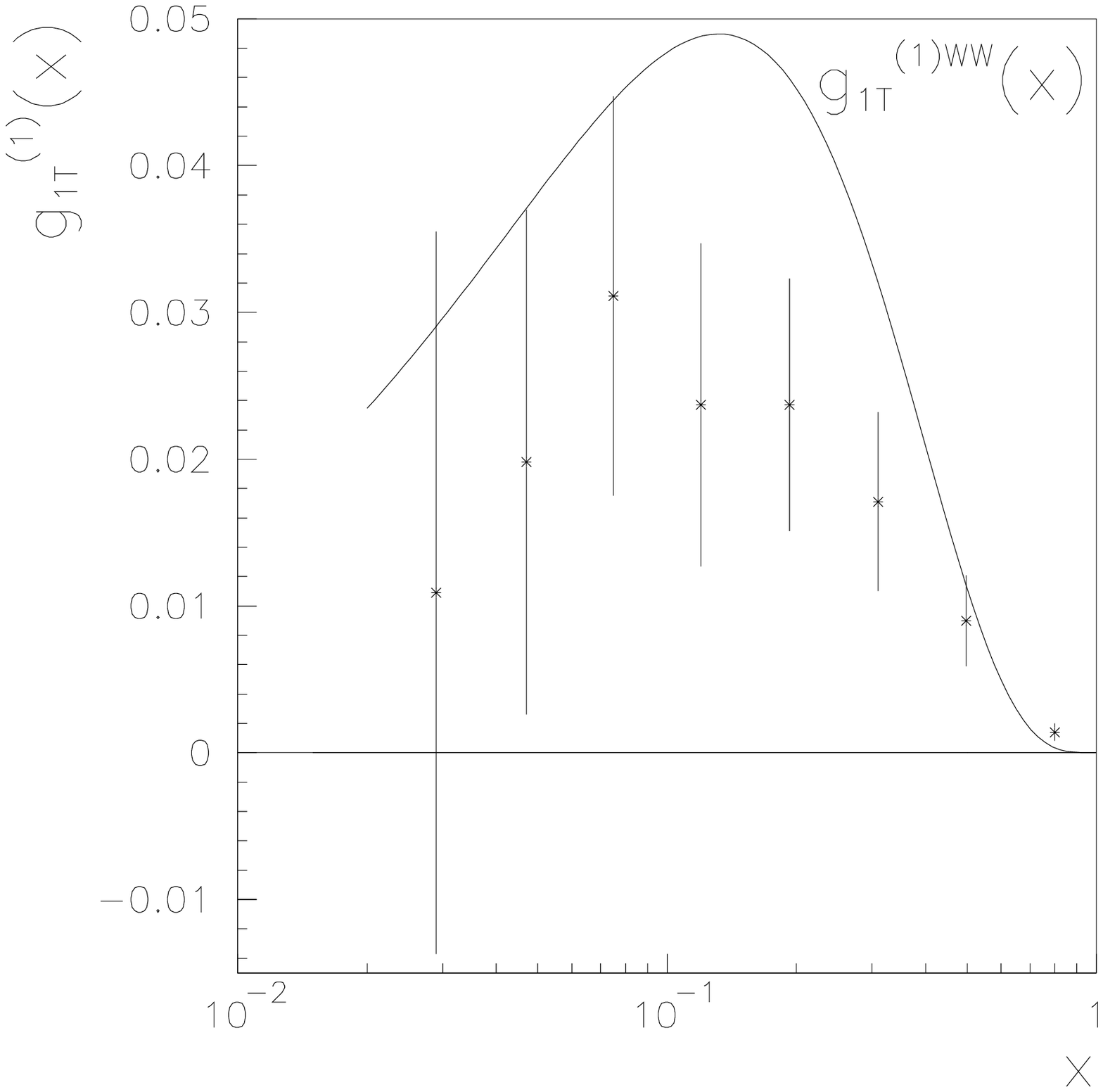,width=6 cm}
{\small Figure 1: The function $g_{1T}^{(1)}(x)$ obtained from the
SLAC E143 data and the WW parametrization.}
\end{wrapfigure}
The relation between $g_2$ and $g_{1T}^{(1)}$ allows an estimate of the
latter using the SLAC E143 data [4] or the Wandzura-Wilczek (WW)
part of $g_2$. This estimate is shown in Fig. 1 and would lead in
$\vec e\vec p \rightarrow e^\prime\pi^+ X$ to an asymmetry proportional to
$g_{1T}^{(1)u}/f_1^u$ which is of the order of $0.05$ [5]. 

A complete analysis of lepton-hadron scattering can be found in ref. [6].
E.g. the functions $h_{1L}^\perp$ and $h_{1T}^\perp$ only appear in combination
with a time-reversal odd fragmentation function $H_1^\perp$ [6,7].
There are several theoretical aspects that have been stepsided here, such as
the inclusion of diagrams dressed with gluons. 
In fact a whole tower of diagrams containing matrix elements with $A^+$ gluon
fields in the target matrix element also contribute at leading order, 
precisely summing up to a gauge link needed to render the nonlocal
matrix element $\Phi$ color gauge invariant. Other gluon contributions are
needed to ensure electromagnetic gauge invariance at order $1/Q$ or they
lead to perturbative QCD corrections.

Summarizing, we stress the fact that inclusion of transverse momenta
of quarks in the formalism of nonlocal matrix elements extends the 
interpretability of structure functions in terms of quark distributions. 
The representation in terms of nonlocal quark fields also provides a
natural link to models for estimating these functions.

Part of this work (R.J. and P.M.) was supported by the foundation for 
Fundamental Research on Matter (FOM) and the Dutch Organization for
Scientific Research (NWO).

\vspace{0.2cm}
\vfill
{\small\begin{description}
\item[1]
D.E. Soper, Phys. Rev. D {\bf 15} (1977) 1141; Phys. Rev. Lett. 43
(1979) 1847;
J.C. Collins and D.E. Soper, Nucl. Phys. B194 (1982) 445;
R.L. Jaffe, Nucl. Phys. B229 (1983) 205
\item[2]
J.P. Ralston and D.E. Soper, Nucl. Phys. B 152 (1979) 109
\item[3]
A.P. Bukhvostov, E.A. Kuraev and L.N. Lipatov, Sov. Phys. JETP 60 (1984) 22
\item[4]
E143 collaboration, K. Abe et al., Phys. Rev. Lett. 76 (1996) 587
\item[5]
A. Kotzinian and P.J. Mulders, Phys. Rev. D54 (1996) 1229
\item[6]
P.J. Mulders and R.D. Tangerman, Nucl. Phys. B461 (1996) 197
\item[7]
R. Jakob and P.J. Mulders, contribution at SPIN96
\end{description}}

\end{document}